\documentclass[%
jmp,%
 sd,%
 amsmath,amssymb,
 reprint,%
]{revtex4-1}

\usepackage{graphicx}
\usepackage{dcolumn}
\usepackage{bm}

\begin{document}

\preprint{AIP/123-QED}

\title[Sample title]{Entropic inequalities for matrix elements of rotation group \\irreducible
representations\footnote{Error!}}

\author{V.I. Man'ko}
\affiliation{P.N. Lebedev Physical Institute, Russian Academy of Sciences,
Leninskii Prospect 53, Moscow 119991, Russia
.}
\author{L.A. Markovich}%
 \email{kimo1@mail.ru}
\affiliation{Moscow Institute of Physics and Technolog\\
 Institutskii Per. 9, Dolgoprudny Moscow Region 141700, Russia
}%

\date{\today}

\begin{abstract}
Using the entropic inequalities for Shannon and Tsallis entropies new inequalities for some classical polynomials are obtained. To this end, an invertible mapping for the irreducible unitary representation of groups $SU(2)$ and $SU(1,1)$ like Jacoby polynomials and Gauss' hypergeometric functions, respectively, are used.
\end{abstract}

\pacs{Valid PACS appear here}
\maketitle

%

\section{\label{sec:1}Introduction}

It is known that matrix elements of some irreducible unitary representation of compact and noncompact groups can be represented
in terms of special functions, e.g. Jacobi and Legendre polynomials \cite{Vil}.
What's more,  any unitary matrix can be associated with a bistochastic matrix. The sums of elements of the bistochastic matrices both in columns
and rows  are equal to one. Hence, it is possible to use the elements of such matrices as probabilities. Symmetry properties provide possibilities to connect the representation
aspects of groups and algebras with properties of special functions (see e.g. \cite{Gromov1,Gromov2,Gromov3,Malkin}).
\par The Shanon entropy characterizes the probability distribution of the random variable which appears as a result of experiment with finite number of outcomes.
For two random variables the joint probability distribution can be obtained.
The distribution is connected with $N=N_1\cdot N_2$ outcomes, where for the first random variable we have $N_1$ results and for the second random variable $N_2$ results.
According to the Sklar's theorem \cite{Nelsen} the joint distribution function and dependence between two random variables determine two marginal distribution functions. For the latter three distributions the Shannon entropies can be calculated. They satisfy the inequality called the subadditivity condition \cite{Lieb}. The entropic inequalities for bipartite systems were used in \cite{Chernega,Mendes} in the
framework of the tomographic probability representation of quantum mechanics to characterize two degrees of quantum correlations in the systems. For the systems without subsystems the latter inequalities were introduced in \cite{Mar2}. However we can apply the subadditivity condition  in all cases, where the set of nonnegative numbers or functions is arisen and the sum of numbers or functions equals to unity.
 For the Li subgroups like $SU(2)$ and $SU(1,1)$ the unitary irreducible representations are well known. For example, they can be represented  in terms of Jacobi, Legendre and Gauss' hypergeometric polynomials, etc. Therefore, we can write some new inequalities for the latter polynomials. The inequalities for the Jacobi and Legendre polynomials in case of the system with the spin $j=3/2$ are introduced in \cite{Mar1}.
\par The aim of our work is to consider the unitary matrices connected with the irreducible representation of the $SU(2)$ and $SU(1,1)$ groups and to construct the new inequalities for such special functions  as Jacobi and Gauss' hypergeometric polynomials from the entropic inequalities.
To this end, the invertible mapping of indices  proposed in \cite{Chernega:2013,ManMan} is used.
\par The paper is organized as follows. In Sec.~\ref{sec:2} the invertible mapping for the finite groups is introduced. In Sec.~\ref{sec:3} we use the latter mapping and the subadditivity condition for the Shannon entropy  to write the new inequalities for the Jacobi polynomials which represent the elements of the $SU(2)$-group. The results are illustrated on examples of the states with the spins $j=3/2$ and $j=2$. Section \ref{sec:4} is dedicated to the use of other invertible mappings and $q$-entropies to write the new inequalities for the special functions. In Sec.~\ref{sec:5} the invertible mapping for the infinite groups is obtained. Using the latter results the new inequalities for the Gauss' hypergeometric polynomials which represent the elements of the $SU(1,1)$-group are written in Sec.~\ref{sec:6}.

\section{\label{sec:2}The invertible mapping for the irreducible unitary representation of the  $SU(2)$-group}

Let $p_1,p_2,\ldots,p_N$  be the set of nonnegative number such that $\sum\limits_{k=1}^{N}p_k=1$, $p_k\geq0$ . The latter numbers can be interpreted as the probability vector
$\overrightarrow{p}$ with  $N$ random components. For the system of qudits with the density matrix $\rho$ the components of the probability vector $\overrightarrow{p}$ are related to the state tomograms \begin{eqnarray*}w(m, u) = <m|u\rho u^{\dagger}|m>,\end{eqnarray*}
 where $u$ is unitary
matrix. The vector $\overrightarrow{m}= (m_1, m_2,\ldots, m_n)$, $m_k = (-j_k, -j_k+1, . . . , j_k)$ is the projection of the $j_k$th spin.
The Shannon entropy associated with the probability vector $\overrightarrow{p}$ is determined by
\begin{eqnarray*}\label{10}H_{p}&=&-\sum\limits_{k}p_{k}\ln p_{k}.
\end{eqnarray*}
In \cite{Mar1} the special invertible mapping of indices was introduced. Namely, if $N$ is an even number than it follows
\begin{eqnarray*}
1&\Leftrightarrow& 11,\quad 2\Leftrightarrow 12,\quad\ldots,\quad\frac{N}{2} \Leftrightarrow 1\frac{N}{2},\\
\frac{N}{2}+1&\Leftrightarrow& 21,\quad \frac{N}{2}+2\Leftrightarrow 22,\quad\ldots,\quad N \Leftrightarrow 2\frac{N}{2}.
\end{eqnarray*}
Hence, the probabilities are given in the form of the matrix $(p_{il})$, $i=1,2$, $l=1,2,\ldots,N/2$ with components
\begin{eqnarray}\label{23_1}
p_1&\Leftrightarrow& p_{11},\quad p_2\Leftrightarrow p_{12},\quad\ldots,\quad p_{\frac{N}{2}} \Leftrightarrow p_{1\frac{N}{2}},\\\nonumber
p_{\frac{N}{2}+1}&\Leftrightarrow& p_{21},\quad p_{\frac{N}{2}+2}\Leftrightarrow p_{22},\quad\ldots,\quad p_{N} \Leftrightarrow p_{2\frac{N}{2}}.
\end{eqnarray}
If $N$ is an odd number than we add a zero component $p_{N+1}=0$ to the $N$-component vector $\overrightarrow{p}$. Then we get the $(N+1)$-component vector $\overrightarrow{p}=(p_1,p_2,\ldots,p_N,p_{N+1})$.
Thus, the invertible map of the indices is the following
\begin{eqnarray*}
1&\Leftrightarrow& 11,\quad\ldots,\quad\frac{N+1}{2} \Leftrightarrow 1\frac{N+1}{2},\\
\frac{N+1}{2}+1&\Leftrightarrow& 21,\quad\ldots,\quad N+1 \Leftrightarrow 2\frac{N+1}{2}.
\end{eqnarray*}
The  probabilities are given in the form of the matrix $(p_{il})$, $i=1,2$, $l=1,2,\ldots,(N+1)/2$ with the components
\begin{eqnarray}\label{24_1}
p_1&\Leftrightarrow& p_{11},\quad p_2\Leftrightarrow p_{12}\quad\ldots,\quad p_{\frac{N+1}{2}} \Leftrightarrow p_{1\frac{N+1}{2}},\\\nonumber
p_{\frac{N+1}{2}+1}&\Leftrightarrow& p_{21},\quad\ldots,\quad p_{N+1} \Leftrightarrow p_{2\frac{N+1}{2}}.
\end{eqnarray}
Let us introduce the unitary matrix $U$ with the matrix elements $u_{ik}$, $i,k=1,2,\ldots, N$ which satisfy the condition
\begin{eqnarray*}\sum\limits_{i=1}^{N}|u_{ik}|^2=\sum\limits_{k=1}^{N}|u_{ik}|^2=1.
\end{eqnarray*}
It is known that the latter $n\times n$ matrix can be associated with the bistochastic matrix $M$, i.e. $|u_{ik}|^2=m_{ik}$.  The sum of numbers both in columns
and rows of bistochastic matrices is equal to one. Let us fix the second index $k$ and introduce the notation $m_{ik}\equiv p_{i}^{(k)}$. Thus, we can map the indices $i$ on the pairs of indices $(\alpha(i),\xi(i))$, $\alpha(i)\in\{1,2\}$, $\xi(i)\in\{1,2,\ldots N/2 ((N+1)/2)\}$ like \eqref{23_1} and \eqref{24_1}.
Using the latter mapping we can rewrite the bistochastic matrix $p_{i}^{(k)}$ in the form $p_{\alpha(i),\xi(i)}^{(k)}$.
\par Since $p_{\alpha(i),\xi(i)}^{(k)}\geq0$ and
$\sum\limits_{\alpha}\sum\limits_{\xi}p_{\alpha(i),\xi(i)}^{(k)}=1$ hold, the values $p_{\alpha(i),\xi(i)}^{(k)}$ can be considered as probabilities. Hence, we can write the Shannon entropy as
\begin{eqnarray}\label{11}\mathcal{H}_{p}(12)&=&-\sum\limits_{\alpha=1}^{2}\sum\limits_{\xi=1}^{\frac{N}{2} (\frac{N+1}{2})}p_{\alpha(i),\xi(i)}^{(k)}\ln p_{\alpha(i),\xi(i)}^{(k)}.
\end{eqnarray}
If we fix one of the indices $\alpha(i)$ or $\xi(i)$ and sum over the unfixed one we can obtain the analog of the marginal distributions
\begin{eqnarray}\label{12}p_{\xi(i)}^{(k)}(1)=\sum\limits_{\alpha=1}^{2}p_{\alpha(i),\xi(i)}^{(k)}, p_{\alpha(i)}^{(k)}(2)=\sum\limits_{\xi=1}^{\frac{N}{2} (\frac{N+1}{2})}p_{\alpha(i),\xi(i)}^{(k)}.
\end{eqnarray}
What's more, we can write two Shannon entropies associated with marginal distributions \eqref{12} as
\begin{eqnarray}\label{13}\mathcal{H}_{p}(1)&=&-\sum\limits_{\xi=1}^{\frac{N}{2} (\frac{N+1}{2})}p_{\xi(i)}^{(k)}(1)\ln p_{\xi(i)}^{(k)}(1),\\\nonumber
\mathcal{H}_{p}(2))&=&-\sum\limits_{\alpha=1}^{2}p_{\alpha(i)}^{(k)}(2)\ln p_{\alpha(i)}^{(k)}(2).
\end{eqnarray}
It is known that these entropies satisfy the subadditivity condition \cite{Lieb} written in the form
of inequality
\begin{eqnarray}\label{14}\mathcal{H}_{p}(1)+\mathcal{H}_{p}(2)\geq\mathcal{H}_{p}(12).
\end{eqnarray}

\section{\label{sec:3} The inequalities for the representation of matrix elements of the $SU(2)$-group}

In this section we deal with the $SU(2)$-group which has the following properties
\begin{eqnarray*}SU(2)=\Bigg\{u=\left(
                                                             \begin{array}{cc}
                                                               a& b \\
                                                            -\overline{b} & \overline{a} \\
                                                             \end{array}
                                                           \right),\quad |a|^2+|b|^2=1,
\Bigg\},\end{eqnarray*}
where $\det u=1$, $a,b\in\mathbf{C}$ and the overline denotes complex conjugation. The complex numbers $a,b$ can be represented using the Euler angels $(\varphi,\theta,\psi)$, $-\pi\leq\varphi\leq\pi$, $0\leq\theta\leq\pi$, $0\leq\psi\leq4\pi$.
 \par From the representation theory it is known that the $SU(2)$-group is generated by $J^{i}=\sigma^{i}/2$, $i\in\{1,2,3\}$. Hence, for the matrix elements of the $SU(2)$ the following parametrization can be used $u=e^{i\psi J^{1}}e^{i\theta J^{2}}e^{i\varphi J^{2}}$. For the latter parametrization the associated $D$-function of the $SU(2)$ reduces to the Wigner $d$-function
\begin{eqnarray*}\label{0}
D_{m'm}^{j}(u)&=&e^{im'\psi}d_{m',m}^{(j)}(\theta)e^{im\varphi}.
\end{eqnarray*}
It is known that the unitary irreducible representations of the rotation group with spins (or $SU(2)$ group) are expressed in terms of Jacobi polynomials \cite{Vil,Landau}.
The squared modules of the matrix elements are determined by
\begin{eqnarray}\label{1}
\Big|d_{m',m}^{(j)}(\theta)\Big|^2&=&S_{m',m}^{(j)}(\theta)\left(P_{m',m}^{(j)}(\theta)\right)^2,
\end{eqnarray}
where the following notation
\begin{eqnarray*}
\quad S_{m',m}^{(j)}(\theta)&=&\frac{(j+m')!(j-m')!}{(j+m)!(j-m)!}\\
&\cdot&\cos\left(\theta/2\right)^{2(m'+m)}\sin\left(\theta/2\right)^{2(m'-m)},
\end{eqnarray*}
is used. $P_{m',m}^{(j)}(\theta)\equiv P_{j-m'}^{(m'-m,m'+m)}(\cos\theta)$ denotes the Jacobi polynomials
\begin{eqnarray*}
P_{n}^{(a,b)}(z)&=& \frac{(-1)^n}{2^nn!}(1-z)^{-a}(1+z)^{-b}\\
&\cdot&\frac{d^n}{dz^n}(1-z)^{a+n}(1+z)^{b+n}.
\end{eqnarray*}
The following relations
\begin{eqnarray}\label{17}
d_{m',m}^{(j)}(\theta)&=&d_{m',m}^{(j)}(\theta), m' + m\geq 0,  m' - m\geq 0,\\\nonumber
d_{m',m}^{(j)}(\theta)&=&d_{-m,-m'}^{(j)}(\theta),m' + m \leq 0,m' - m \geq 0,\\ \nonumber
d_{m',m}^{(j)}(\theta)&=&(-1)^{m'-m} d_{m,m'}^{(j)}(\theta),\\ \nonumber
&&m' + m \geq 0,m' - m \leq 0,\\\nonumber
d_{m',m}^{(j)}(\theta)&=&(-1)^{m'-m} d_{-m',-m}^{(j)}(\theta),\\ \nonumber
&& m' + m \leq 0,  m' - m \leq 0.\nonumber
\end{eqnarray}
hold \cite{Andrews}.
We shall apply the inequalities for probabilities expressed in terms of Shannon entropies  \cite{Shannon} to the matrix elements \eqref{1}.
The point is that one has $\Big|d_{m'm}^{(j)}(\theta)\Big|^2\geq0$ and
\begin{eqnarray*}\label{2}
\sum\limits_{m'=-j}^{j}\Big|d_{m',m}^{(j)}(\theta)\Big|^2&=&\sum\limits_{m=-j}^{j}\Big|d_{m',m}^{(j)}(\theta)\Big|^2=1.
\end{eqnarray*}
Thus, the values $\Big|d_{m'm}^{(j)}(\theta)\Big|^2$ can be considered as probabilities.
We denote these probabilities as
\begin{eqnarray}\label{3}
p_{m'm}^{(j)}(\theta)&=&\Big|d_{m',m}^{(j)}(\theta)\Big|^2.
\end{eqnarray}
We shall use the map of numbers $m'$ and $m$ onto the numbers $1,2,\ldots,N$, $N=2j+1$ using the following rule
$-j\Rightarrow 1,\quad -j+1\Rightarrow 2,\quad\ldots,\quad j \Rightarrow N.$
Thus, we can study the relation which can be obtained by considering the probability vector $\overrightarrow{p}=(p_1,p_2,\ldots,p_N)$, where
$\sum\limits_{k=1}^{N}p_k=1$, $p_k\geq0$ hold. Hence, similarly to Sec. \ref{sec:1} we can denote \eqref{3} as $\widetilde{m}_{ik}$, $i,k\in\{1,2,\ldots N\}$. Fixing the index $k$ and mapping the index $i$ onto pairs $(\alpha(i),\beta(i))$ as in \eqref{23_1} or \eqref{24_1} we can write the entropy \eqref{11} of the whole system  in terms of $\widetilde{m}_{ik}\equiv\widetilde{p}_{\alpha(i),\xi(i)}^{(k)}$, i.e.
\begin{eqnarray*}\widetilde{\mathcal{H}}_{m}(12)&=&-\sum\limits_{\alpha=1}^{2}\sum\limits_{\xi=1}^{\frac{N}{2} (\frac{N+1}{2})}\widetilde{p}_{\alpha(i),\xi(i)}^{(k)}\ln \widetilde{p}_{\alpha(i),\xi(i)}^{(k)}\\
&=&-\sum\limits_{\xi=1}^{\frac{N}{2} (\frac{N+1}{2})}\left(\widetilde{p}_{1,\xi(i)}^{(k)}\ln \widetilde{p}_{1,\xi(i)}^{(k)}+\widetilde{p}_{2,\xi(i)}^{(k)}\ln \widetilde{p}_{2,\xi(i)}^{(k)}\right)\\
&=&-\sum\limits_{i=1}^{N(N+1)}\widetilde{m}_{ik}\ln \widetilde{m}_{ik}\\
&=&-\sum\limits_{m'=-j}^{j}\Big|d_{m',m}^{(j)}(\theta)\Big|^2\ln \Big|d_{m',m}^{(j)}(\theta)\Big|^2.
\end{eqnarray*}
Analogically, the partial entropies \eqref{13} can be written  as
\begin{eqnarray*}
\widetilde{\mathcal{H}}_{m}(1)&=&-\sum\limits_{\xi=1}^{\frac{N}{2} (\frac{N+1}{2})}\widetilde{p}_{\xi(i)}^{(k)}(1)\ln \widetilde{p}_{\xi(i)}^{(k)}(1)\\
&=&-\sum\limits_{m_1'=-j}^{-\frac{1}{2}(0)}\sum\limits_{m_2'=\frac{1}{2}(1)}^{j}\left(\Big|d_{m_1'm}^{(j)}(\theta)\Big|^2+\Big|d_{m_2'm}^{(j)}(\theta)\Big|^2\right)\\
&\cdot&\ln \left(\Big|d_{m_1'm}^{(j)}(\theta)\Big|^2+\Big|d_{m_2'm}^{(j)}(\theta)\Big|^2\right),
\end{eqnarray*}
\begin{eqnarray*}
&&\widetilde{\mathcal{H}}_{m}(2)=-\sum\limits_{\alpha=1}^{2}\widetilde{p}_{\alpha(i)}^{(k)}(2)\ln \widetilde{p}_{\alpha(i)}^{(k)}(2)\\
&=&-\sum\limits_{m'=-j}^{-\frac{1}{2}(0)}\Big|d_{m',m}^{(j)}(\theta)\Big|^2\ln\left(\sum\limits_{m'=-j}^{-\frac{1}{2}(0)}\Big|d_{m',m}^{(j)}(\theta)\Big|^2\right)\\
&-&\sum\limits_{m'=\frac{1}{2}(1)}^{j}\Big|d_{m',m}^{(j)}(\theta)\Big|^2\ln\left(\sum\limits_{m'=\frac{1}{2}(1)}^{j}\Big|d_{m',m}^{(j)}(\theta)\Big|^2\right).
\end{eqnarray*}
The subadditivity condition \eqref{14} for the matrix from the $SU(2)$-group is the following
\begin{widetext}
       \begin{eqnarray*}\label{15}&-&{\sum\limits_{m_1'=-j}^{-\frac{1}{2}(0)}\sum\limits_{m_2'=\frac{1}{2}(1)}^{j}
\Bigg(S_{m_1',m}^{(j)}(\theta)P_{m_1',m}^{(j)}(\theta)^2+S_{m_2',m}^{(j)}(\theta)P_{m_2',m}^{(j)}(\theta)^2\Bigg)}\ln \left(S_{m_1',m}^{(j)}(\theta)P_{m_1',m}^{(j)}(\theta)^2
+S_{m_2',m}^{(j)}(\theta)P_{m_2',m}^{(j)}(\theta)^2\right)\\\nonumber
&-&\!\!\sum\limits_{m'=-j}^{-\frac{1}{2}(0)}S_{m',m}^{(j)}(\theta)P_{m',m}^{(j)}(\theta)^2
\ln\left(\sum\limits_{m'=-j}^{-\frac{1}{2}(0)}S_{m',m}^{(j)}(\theta)P_{m',m}^{(j)}(\theta)^2\right)-\!\!\!\!\sum\limits_{m'=\frac{1}{2}(1)}^{j}\!\!\!\!S_{m',m}^{(j)}(\theta)P_{m',m}^{(j)}(\theta)^2\ln \left(\sum\limits_{m'=\frac{1}{2}(1)}^{j}S_{m',m}^{(j)}(\theta)P_{m',m}^{(j)}(\theta)^2\right)\\\nonumber
&\geq&-\sum\limits_{m'=-j}^{j}S_{m',m}^{(j)}(\theta)P_{m',m}^{(j)}(\theta)^2\Bigg(\ln\left(S_{m',m}^{(j)}(\theta)P_{m',m}^{(j)}(\theta)^2\right)\Bigg).
\end{eqnarray*}
\end{widetext}
The resulted inequality can be interpreted as the new inequality for the Jacoby polynomials.

\subsection{Examples of systems with spins $j=3/2$ and $j=2$}

Let us consider the state with the spin $j=3/2$. As an example we take $m=3/2$. Hence, the partial entropies are determined by
\begin{eqnarray*}\widetilde{\mathcal{H}}_{3/2}(12)&=&-\sum\limits_{t=1}^{4}(p_t(\theta)\ln(p_t(\theta)),\\
\widetilde{\mathcal{H}}_{3/2}(1)&=&-((p_3(\theta)+p_1(\theta))\ln((p_3(\theta)+p_1(\theta)))\\
&+&(p_4(\theta)+p_1(\theta))\ln((p_4(\theta)+p_1(\theta)))\\
&+&(p_3(\theta)+p_2(\theta))\ln((p_3(\theta)+p_2(\theta)))\\
&+&(p_4(\theta)+p_2(\theta))\ln((p_4(\theta)+p_2(\theta)))),\\
\widetilde{\mathcal{H}}_{3/2}(2)&=&-((p_3(\theta)+p_4(\theta))\ln((p_3(\theta)+p_4(\theta)))\\
&+&(p_2(\theta)+p_1(\theta))\ln((p_2(\theta)+p_1(\theta)))),
\end{eqnarray*}
where  we denote
\begin{eqnarray*}p_1(\theta)&=&(\cos\theta+1)^3/8,\\ p_2(\theta)&=&3\sin^2(\theta/2)(\sin^2(\theta/2)-1)^2,\\
p_3(\theta)&=&3(\cos\theta-1)^3(\cos\theta+1)/8,\\ p_4(\theta)&=&-(\cos\theta-1)^3/8.
\end{eqnarray*}
Then we obtain the subadditivity condition \eqref{15}.
\par
\begin{figure}[ht]
\includegraphics[width=1\linewidth]{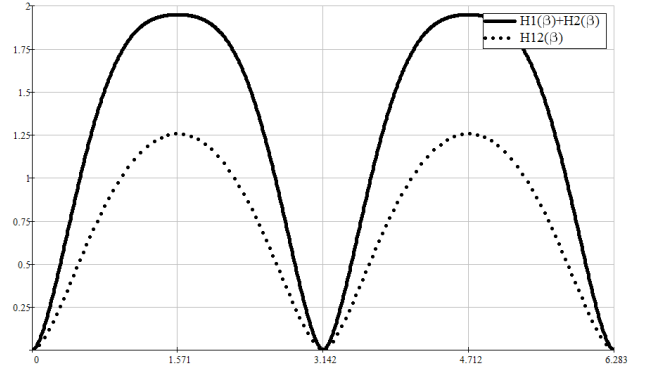}
\vspace{-4mm}
\caption{\label{fig:1}The left hand side (black line) and the right hand side (dotted line) of the subadditivity condition \eqref{15} for the system with the spin $j=3/2$.}
\end{figure}

\begin{figure}[ht]
\includegraphics[width=1\linewidth]{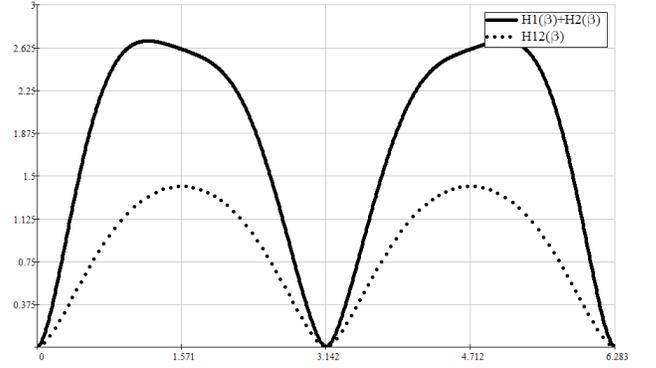}
\vspace{-4mm}
\caption{\label{fig:2}The left hand side (black line) and the right hand side (dotted line) of the subadditivity condition \eqref{15} for the system with the spin $j=2$.}
\end{figure}

 Analogically, for the state with the spin $j=2$ and $m=2$ we can write the partial  entropies as
\begin{eqnarray*}\widetilde{\mathcal{H}}_{2}(12)&=&-\sum\limits_{i=1}^{5}(t_i(\theta)\ln(t_i(\theta)),\\
\widetilde{\mathcal{H}}_{2}(1)&=&-((t_1(\theta)+t_5(\theta))\ln((t_1(\theta)+t_5(\theta)))\\
&+&(t_2(\theta)+t_5(\theta))\ln(t_2(\theta)+t_5(\theta)))\\
&+&(t_1(\theta)+t_4(\theta))\ln(t_1(\theta)+t_4(\theta)))\\
&+&(t_1(\theta)+t_3(\theta))\ln(t_1(\theta)+t_3(\theta))\\
&+&(t_2(\theta)+t_3(\theta))\ln(t_2(\theta)+t_3(\theta)))\\
&+&(t_2(\theta)+t_4(\theta))\ln(t_2(\theta)+t_4(\theta)),\\
\widetilde{\mathcal{H}}_{2}(2)&=&-((t_2(\theta)+t_1(\theta))\ln((t_2(\theta)+t_1(\theta)))\\
&+&(t_1(\theta)+t_4(\theta)+t_5(\theta))\ln(t_1(\theta)+t_4(\theta)+t_5(\theta))),
\end{eqnarray*}
where we denote
\begin{eqnarray*}t_1(\theta)&=&(\cos\theta+1)^4/16,\\ t_2(\theta)&=&4\cos(\theta/2)^6(1-\cos(\theta/2)^2),\\
t_3(\theta)&=&3\sin\theta^4/8,\quad t_4(\theta)=4\sin(\theta/2)^6(1-\sin(\theta/2)^2),\\ t_5(\theta)&=&(\cos\theta-1)^4/16.
\end{eqnarray*}
Hence, we can write the subadditivity condition \eqref{15} for the system with the spin $j=2$.
The obtained results for various angles $\beta$ are shown in Fig.~\ref{fig:1} and \ref{fig:2}.
The sum of the entropies $\widetilde{\mathcal{H}}_{m}(1)$, $\widetilde{\mathcal{H}}_{m}(2)$ is shown by the black lines and  the entropy of the whole system $\widetilde{\mathcal{H}}_{m}(12)$ by the dotted lines.
Needless to say that the sum of the partial entropies is higher than the entropy of the whole system. The equality is reached only in the points $\beta=\{0,\pi,2\pi\}$.

\section{\label{sec:4} Examples of other invertible mappings and entropies}

Using various invertible mappings we can get many inequalities for the special functions. To this end, let us introduce the following mapping
\begin{eqnarray*}
1&\Leftrightarrow& 11,\quad 2\Leftrightarrow 12,\quad \quad\ldots,\quad N_1 \Leftrightarrow N_1 1,\\
N_1+1&\Leftrightarrow&  21,\quad\ldots,\quad N \Leftrightarrow N_1N_2.
\end{eqnarray*}
It means that we use the invertible map of natural numbers $1,2,\ldots,N$ onto pairs of integers $(i,k)$, $i\in\{1,2,\ldots,N_1\}$, $k\in\{1,2,\ldots,N_2\}$.
Therefore, using the latter indices we can write the Shannon entropies as following
\begin{eqnarray}\label{22}\mathcal{H}_{p}(12)&=&-\sum\limits_{\alpha=1}^{N_1}\sum\limits_{\xi=1}^{N_2}p_{\alpha(i),\xi(i)}^{(k)}\ln p_{\alpha(i),\xi(i)}^{(k)},
\end{eqnarray}
\begin{eqnarray*}\label{23}\mathcal{H}_{p}(1)&=&-\sum\limits_{\xi=1}^{N_2}p_{\xi(i)}^{(k)}(1)\ln p_{\xi(i)}^{(k)}(1),\\
\mathcal{H}_{p}(2)&=&-\sum\limits_{\alpha=1}^{N_1}p_{\alpha(i)}^{(k)}(2)\ln p_{\alpha(i)}^{(k)}(2).
\end{eqnarray*}
Needless to say that the latter entropies satisfy the subadditivity condition. Hence, similarly to Sec. \ref{sec:2} we can use it to write the new inequalities for the special functions.
\par
In \cite{Chernega:2013,ManMan} the following mapping has been introduced. The probability vector $p$ introduced in Sec. \ref{sec:2} with the components $p_i$, $i \in\{ 1, 2, \ldots , N\}$ is mapped onto the table of numbers with three indices $\Pi_{kjl}$, $k \in\{1, 2,\ldots, N_1\}$, $j\in\{ 1, 2,\ldots, N_2\}$, $l\in\{1, 2,\ldots, N_3\}$.
Hence, we can consider that the system has three subsystems with the three random variables and the joint probability distribution  describing the results of measurement of
the random variables is related to the nonnegative numbers. The nonnegative numbers determine the marginal
probability distributions. Hence, we can do the same procedure as in  Sec. \ref{sec:2} and write the new inequalities for the special functions.
\par However, instead of using the Shannon entropy we can select other entropies, for example the $q$-deformed entropies like Tsallis  and R\'{e}nyi \cite{Renyi,Tsallis}
\begin{eqnarray*}\label{36}S_{q}^{T}&=&\left(\sum\limits_{i=1}^{\infty}p_i^q-1\right)/(1-q),\\
 S_q^{R} &=&\ln\left(\sum\limits_{i=1}^{\infty}p_i^q\right)/(1-q).
\end{eqnarray*}
These entropies being functions of an extra parameter contain more detailed information on properties
of density matrices of the qudit states and the qudit subsystem states. The Tsallis  entropy of the bipartite qudit system was shown to satisfy the generalized subadditivity condition \cite{Audenaert,Petz}. This condition is the inequality available for Tsallis entropy of the bipartite system state and Tsallis entropies of two subsystem states.
\par Hence, we can write the Tsallis entropies for the mapping obtained in Sec. \ref{sec:3}
\begin{eqnarray*}S_{q}^{T}(1,2)&=&\left(\sum\limits_{\alpha=1}^{2}\sum\limits_{\xi=1}^{\frac{N}{2} (\frac{N+1}{2})}\left(p_{\alpha(i),\xi(i)}^{(k)}\right)^q-1\right)/(1-q),
\end{eqnarray*}
\begin{eqnarray*}S_{q}^{T}(1)&=&\left(\sum\limits_{\xi=1}^{\frac{N}{2} (\frac{N+1}{2})}\left(p_{\xi(i)}^{(k)}(1)\right)^q-1\right)/(1-q),\\ S_{q}^{T}(2)&=&\left(\sum\limits_{\alpha=1}^{2}\left(p_{\alpha(i)}^{(k)}(2)\right)^q-1\right)/(1-q).
\end{eqnarray*}
Next, using the subadditivity of the Tsallis entropy we can write
\begin{eqnarray*}&&\sum\limits_{\xi=1}^{\frac{N}{2} (\frac{N+1}{2})}\left(p_{\xi(i)}^{(k)}(1)\right)^q+\sum\limits_{\alpha=1}^{2}\left(p_{\alpha(i)}^{(k)}(2)\right)^q-1\geq\\
&\geq&\sum\limits_{\alpha=1}^{2}\sum\limits_{\xi=1}^{\frac{N}{2} (\frac{N+1}{2})}\left(p_{\alpha(i),\xi(i)}^{(k)}\right)^q.\end{eqnarray*}
Substituting the polynomial \eqref{1} in the latter inequality  we obtain the new inequality for the Jacoby polynomials
\begin{widetext}
       \begin{eqnarray*}&&{\sum\limits_{m_1'=-j}^{-\frac{1}{2}(0)}\sum\limits_{m_2'=\frac{1}{2}(1)}^{j}
\Bigg(S_{m_1',m}^{(j)}(\theta)\left(P_{m_1',m}^{(j)}(\theta)\right)^2+S_{m_2',m}^{(j)}(\theta)\left(P_{m_2',m}^{(j)}(\theta)\right)^2\Bigg)^q}\\\nonumber
&+&\sum\limits_{m'=-j}^{-\frac{1}{2}(0)}\left(S_{m',m}^{(j)}(\theta)\left(P_{m',m}^{(j)}(\theta)\right)^2\right)^q+\sum\limits_{m'=\frac{1}{2}(1)}^{j}\left(S_{m',m}^{(j)}(\theta)\left(P_{m',m}^{(j)}(\theta)\right)^2\right)^q\geq\sum\limits_{m'=-j}^{j}\left(S_{m',m}^{(j)}(\theta)\left(P_{m',m}^{(j)}(\theta)\right)^2\right)^q.
\end{eqnarray*}
\end{widetext}
Needless to say that we can write the variety of such inequalities using various mappings and entropies.

\section{\label{sec:5} The invertible mapping for the irreducible unitary representations of the $SU(1,1)$-groups}

Let us consider the infinite sets of numbers $m',m\in\{-j,-j+1,-j+2,\ldots\}$, $m',m\in\{j,j-1,j-2,\ldots\}$, $m',m\in\{0,\pm1,\pm2,\ldots\}$ or $m',m\in\{\pm1/2,\pm3/2,\ldots\}$.
We shall use the map of the numbers $m'$ and $m$ onto the numbers $1,2,\ldots$ using the following rules
\begin{eqnarray*}\label{25}
&&-j\Rightarrow 1,\quad -j+1\Rightarrow 2,\quad-j+2\Rightarrow 3,\\\nonumber
&& -j+3\Rightarrow 4,\quad-j+4\Rightarrow 5,\quad -j+5\Rightarrow 6,\ldots\\\nonumber
&&j\Rightarrow 1,\quad j-1\Rightarrow 2,\quad j-2\Rightarrow 3,\\\nonumber
&& j-3\Rightarrow 4,\quad j-4\Rightarrow 5,\quad j-5\Rightarrow 6,\ldots,\nonumber
\end{eqnarray*}
\begin{eqnarray*}\label{26}&&0\Rightarrow 1,\quad 1\Rightarrow 2,\quad -1\Rightarrow 3,\\\nonumber
&&2\Rightarrow 4,\quad-2\Rightarrow 5,\quad 3\Rightarrow 6,\quad -3\Rightarrow 7,\ldots,\\\nonumber
&&-1/2\Rightarrow 1,\quad 1/2\Rightarrow 2,\quad -3/2\Rightarrow 3,\\\nonumber
&&3/2\Rightarrow 4,-5/2\Rightarrow 5,\quad 5/2\Rightarrow 6,\ldots.\nonumber
\end{eqnarray*}
Thus, we can consider the probability vector $\overrightarrow{p}=(p_1,p_2,p_3,\ldots)$, where $\sum\limits_{k=1}^{\infty}p_k=1$, $p_k\geq0$ hold.
\par Let us introduce the  diagonal matrix $\rho_{12}$ with the elements $p_{m'}^{(m)}$
 \begin{eqnarray*}
\rho_{12}&=&\left(
              \begin{array}{cccc}
                p_1^{(m)} & 0 & 0 & \cdots \\
                0& p_2^{(m)}& 0& \cdots \\
                0 & 0 & p_3^{(m)} & \cdots \\
                \vdots & \vdots& \vdots & \ddots \\
              \end{array}
            \right).
\end{eqnarray*}
Let us partition the latter matrix into block matrices of the size $2\times 2$. Hence, we can construct two new matrices using the following rules
 \begin{eqnarray*}
\rho_{1}&=&\left(
              \begin{array}{cccc}
                p_1^{(m)} + p_2^{(m)}& 0 & 0 & \cdots \\
                0& p_3^{(m)} +p_4^{(m)}& 0& \cdots \\
                0 & 0 & p_5^{(m)}+p_6^{(m)}& \cdots \\
                \vdots & \vdots& \vdots & \ddots \\
              \end{array}
            \right),
\end{eqnarray*}
 \begin{eqnarray*}
\rho_{2}\!&=&\!\!\left(
            \begin{array}{cc}
              p_1^{(m)} & 0 \\
              0 & p_2^{(m)} \\
            \end{array}
          \right)\!+\!\left(
            \begin{array}{cc}
              p_3^{(m)} & 0 \\
              0 & p_4^{(m)} \\
            \end{array}
          \right)\!+\!\left(
            \begin{array}{cc}
              p_5^{(m)} & 0 \\
              0 & p_6^{(m)} \\
            \end{array}
          \right)\!+\!\ldots\\
&=&\!\!
\left(
  \begin{array}{cc}
    p_1^{(m)}+p_3^{(m)}+p_5^{(m)}+\ldots& 0 \\
    0 & p_2^{(m)}+p_4^{(m)}+p_6^{(m)}\ldots \\
  \end{array}
\right).
\end{eqnarray*}
Hence, the Shannon entropy can be written as
\begin{eqnarray*}\mathcal{H}(12)&=&-\sum\limits_{k=1}^{\infty}p_k^{(m)}\ln p_k^{(m)}.
\end{eqnarray*}
The Shannon entropies for the subsystems are the following
\begin{eqnarray*}\mathcal{H}(1)&=&-\sum\limits_{k=2i+1}^{\infty}(p_k^{(m)}+p_{k+1}^{(m)})\ln (p_k^{(m)}+p_{k+1}^{(m)}), \quad i\in\mathbb{Z},
\end{eqnarray*}
\begin{eqnarray*}
\mathcal{H}(2)&=&-\left(\sum\limits_{k=2i+1}^{\infty}p_k^{(m)}\right)\ln \left(\sum\limits_{k=2i+1}^{\infty}p_k^{(m)}\right)\\\nonumber
&-&\left(\sum\limits_{k=2t}^{\infty}p_k^{(m)}\right)\ln \left(\sum\limits_{k=2t}^{\infty}p_k^{(m)}\right),\quad t\in\mathbb{N}.
\end{eqnarray*}
What's more, we can write the subadditivity condition as
\begin{eqnarray}\label{27}&&-\sum\limits_{k=2i+1}^{\infty}(p_k^{(m)}+p_{k+1}^{(m)})\ln (p_k^{(m)}+p_{k+1}^{(m)})\\\nonumber
&-&\left(\sum\limits_{k=2i+1}^{\infty}p_k^{(m)}\right)\ln \left(\sum\limits_{k=2i+1}^{\infty}p_k^{(m)}\right)\\\nonumber
&-&\left(\sum\limits_{k=2t}^{\infty}p_k^{(m)}\right)\ln \left(\sum\limits_{k=2t}^{\infty}p_k^{(m)}\right)
\geq -\sum\limits_{k=1}^{\infty}p_k^{(m)}\ln p_k^{(m)}.
\end{eqnarray}

\section{\label{sec:6} The inequalities for the representation of matrix elements of the $SU(1,1)$-group}

Let us consider the $SU(1,1)$-group  which has a representation as the group of complex matrices
\begin{eqnarray*}SU(1,1)=\Bigg\{u=\left(
                                                             \begin{array}{cc}
                                                               a& b \\
                                                            \overline{b} & \overline{a} \\
                                                             \end{array}
                                                           \right),\quad |a|^2-|b|^2=1,
\Bigg\},\end{eqnarray*}
where $ \det u=1$, $ a,b\in\mathbf{C}$. The $SU(1,1)$ is generated by $J^{3},K^{1},K^{2}$, $K^{i}=i\sigma^{i}/2$, $i\in\{1,2,3\}$. Unitary irreps of $SU(1,1)$ have two classes, the discrete and the continuous series. For the discrete series the spin $j=-k/2$, $k\in \mathbb{N}$ and the states $|jm>$ have the eigenvalues $m\in\{-j,-j+1,-j+2,\ldots\}$ and $m\in\{j,j-1,j-2,\ldots\}$. For the continuous series the spin is $j=-1/2+is$, $0<s<\infty$ and  $m\in\{0,\pm1,\pm2,\ldots\}$ or $m\in\{\pm1/2,\pm3/2,\ldots\}$.
\par If we consider the case of $SU(1,1)$ elements parameterized as in \cite{Conrady}
\begin{eqnarray*}\label{19}&&u=e^{i\psi J^{3}}e^{it K^{2}}e^{i\varphi J^{3}}, \\\nonumber
 &&0\leq\psi\leq4\pi,\quad 0\leq t<\infty,\quad -\pi\leq\varphi\leq\pi.\end{eqnarray*}
For both the discrete and continuous series the $D$-function is
\begin{eqnarray*}\label{01}
D_{m'm}^{j}(\upsilon)&=&e^{im'\psi}b_{m',m}^{(j)}(t)e^{im\varphi},
\end{eqnarray*}
where $b_{m',m}^{(j)}(t)$ is the Bargmann $b$-function \cite{Bargmann}, the analog of the Wigner $d$-function \eqref{1} in the group $SU(1,1)$.
It is connected with the $d$-function as follows
\begin{eqnarray}\label{33}b^{j}_{m'm}(t)&=&\sqrt{(-1)^{m'-m}}d^{j}_{m'm}(it).\end{eqnarray}
For the case when $m'+m\geq0$, $m'-m\geq0$ the explicit form of the latter is
\begin{eqnarray}\label{35}
b^{j}_{m'm}(t)&=&N_{m'm}^{j}F_{m'm}^{j}(z(it)),
\end{eqnarray}
where $z(it)=(1-\cos it)/2$, the normalization factor is
\begin{eqnarray*}N_{m'm}^{j}&=&\left(\frac{\Gamma(m'+j+1)\Gamma(m'-j)}{\Gamma(m+j+1)\Gamma(m-j)}\right)^{1/2}
\end{eqnarray*}
and
\begin{eqnarray*}&&F_{m'm}^{j}(z(it))=(1-z(it))^{(m'+m)/2}z(it)^{(m'-m)/2}\\
&\cdot&{}_2F_{1}(-j+m',j+m'+1,m'-m+1;z(it))
\end{eqnarray*}
where ${}_2F_{1}$ denotes the Gauss' hypergeometric function with $F_{m'm}^{j}(z(it))=F_{m'm}^{-j-1}(z(it))$. For the
other three variants of $m',m$ we use \eqref{17}.
\par Let us consider the class of $SU(1,1)$ elements parameterized as in  \cite{Conrady,Lindblad}
\begin{eqnarray*}\label{21}u=e^{i\psi J^{3}}e^{it K^{2}}e^{ir K^{1}},0\leq\psi\leq4\pi, 0\leq t,r<\infty.\end{eqnarray*}
In this mixed basis  the left state belongs to the discrete basis ($m'\in\{-j,-j+1,-j+2,\ldots\}$ or $m'\in\{j,j-1,j-2,\ldots\}$) and the right state to the continuous basis. The $D$-function is the following
\begin{eqnarray*}
D_{m'm}^{j}(\upsilon)&=&e^{im'\varphi}c_{m'm}^{(j)}(t)e^{imr}.
\end{eqnarray*}
The function $c_{m'm}^{(j)}(t)$ is
\begin{eqnarray*}\label{34}c_{m'm}^{(j)}(t)&=&N_{m'm}^{j}F^{j}_{-m',-im}(z(-t)),\quad m'\geq-j,\\\nonumber
c_{m'm}^{(j)}(t)&=&g_{-m'm}^{(j)}(-t),\quad m'\leq j,
\end{eqnarray*}
where $z(t)=(1-i\sinh t)/2$ and
\begin{eqnarray*}N_{m'm}&=&\sqrt{2}2^{-j-2}S_{m'}^{j}R_{m'm}^{j}/\pi,
\end{eqnarray*}
\begin{eqnarray*}
 S_{m'}^{j}&=&\sqrt{\Gamma(m'-j)\Gamma(m'+j+1)}/\Gamma(m'+j+1),
 \end{eqnarray*}
 \begin{eqnarray*}
R_{m'm}^{j}&=&\frac{\Gamma(j+1+im)\Gamma(\frac{-j-im}{2})\Gamma(\frac{-j+1+im}{2})}{\Gamma(m'-j)\Gamma(-m'+1+im)}.
\end{eqnarray*}
For the continuous series the $D$-function is
\begin{eqnarray*}
D_{m'm\sigma}^{j}(\upsilon)&=&e^{im'\varphi}l_{m'm\sigma}^{(j)}(t)e^{imr},
\end{eqnarray*}
where
\begin{eqnarray}\label{32}l_{m'm\sigma}^{(j)}(t)&=&S_{m'}^{j}\Bigg(T_{m'm\sigma}^{j}F_{m',-im}^{j}(z(t))\\\nonumber
&-&(-1)^\sigma T_{-m'm\sigma}^{j}F_{-m',-im}^{j}(z(-t))\Bigg).
\end{eqnarray}
The following notation is used
\begin{eqnarray*}T_{m'm\sigma}^{j}&=&\frac{2^{j-1}}{i^\sigma\sin(\pi(-j+\sigma-im)/2)}\\
&\cdot&\frac{\Gamma(-j+im)}{\Gamma(-m'-j)\Gamma(m'+1+im)}.
\end{eqnarray*}
\par Using the results obtained in Sec. \ref{sec:4} let us write the new inequalities for the Gauss' hypergeometric function. Hereafter, the unitary matrix $U$  with the matrix elements $u_{m'm}^{j}$ is from the group $SU(1,1)$.
If the series is discrete positive the indexes are  $m'\in\{-j,-j+1,-j+2,\ldots\}$ and using \eqref{27} we can write
\begin{widetext}\begin{eqnarray}\label{28}&-&{\sum\limits_{m'=-j+2i+1}^{\infty}(|u_{m',m}|^2+|u_{m'+1,m}|^2)\ln (|u_{m',m}|^2+|u_{m'+1,m}|^2)}-\left(\sum\limits_{m'=-j+2t}^{\infty}|u_{m',m}|^2\right)\ln \left(\sum\limits_{m'=-j+2t}^{\infty}|u_{m',m}|^2\right)\nonumber\\
&-&\left(\sum\limits_{m'=-j+2i+1}^{\infty}|u_{m',m}|^2\right)\ln \left(\sum\limits_{m'=-j+2i+1}^{\infty}|u_{m',m}|^2\right)\geq -\sum\limits_{m'=-j}^{\infty}|u_{m',m}|^2\ln |u_{m',m}|^2
\end{eqnarray}
      \end{widetext}
and for the discrete negative series the indexes are $m'\in\{j,j-1,j-2,\ldots\}$ and the inequality is
\begin{widetext}
\begin{eqnarray*}\label{29}&-&{\sum\limits_{m'=j-2i-1}^{\infty}(|u_{m',m}|^2+|u_{m'+1,m}|^2)\ln (|u_{m',m}|^2+|u_{m'+1,m}|^2)}-\left(\sum\limits_{m'=j-2t}^{\infty}|u_{m',m}|^2\right)\ln \left(\sum\limits_{m'=j-2t}^{\infty}|u_{m',m}|^2\right)\nonumber\\
&-&\left(\sum\limits_{m'=j-2i-1}^{\infty}|u_{m',m}|^2\right)\ln \left(\sum\limits_{m'=j-2i-1}^{\infty}|u_{m',m}|^2\right)\geq -\sum\limits_{m'=j}^{-\infty}|u_{m',m}|^2\ln |u_{m',m}|^2,
\end{eqnarray*}
      \end{widetext}
where instead of $u_{m'm}^{j}$ we must substitute \eqref{33} and \eqref{34}. For example, if the series is discrete negative and the matrix elements are defined in \eqref{35} we can write the following inequality for the Gauss' hypergeometric function
\begin{widetext}
\begin{eqnarray*}&-&{\sum\limits_{m'=-j+2i+1}^{\infty}(|N_{m'm}^{j}F_{m'm}^{j}(z)|^2+|N_{m'+1m}^{j}F_{m'+1,m}^{j}(z)|^2)}\ln (|N_{m'm}^{j}F_{m'm}^{j}(z)|^2+|N_{m'+1,m}^{j}F_{m'+1,m}^{j}(z)|^2)\\\nonumber
&-&\left(\sum\limits_{m'=-j+2t}^{\infty}|N_{m'm}^{j}F_{m'm}^{j}(z)|^2\right)\ln \left(\sum\limits_{m'=-j+2t}^{\infty}|N_{m'm}^{j}F_{m'm}^{j}(z)|^2\right)
-\left(\sum\limits_{m'=-j+2i+1}^{\infty}|N_{m'm}^{j}F_{m'm}^{j}(z)|^2\right)\\\nonumber
&\cdot&\ln \left(\sum\limits_{m'=-j+2i+1}^{\infty}|N_{m'm}^{j}F_{m'm}^{j}(z)|^2\right)
\geq -\sum\limits_{m'=-j}^{\infty}|N_{m'm}^{j}F_{m'm}^{j}(z)|^2\ln |N_{m'm}^{j}F_{m'm}^{j}(z)|^2.
\end{eqnarray*}
      \end{widetext}
For the continuous series the matrix elements are defined in \eqref{32} and the inequalities are
\begin{widetext}
       \begin{eqnarray*}\label{30}&-&{\sum\limits_{m'=0}^{-\infty}(|l_{m'm\sigma}^{(j)}(t)|^2+|l_{m'+1,m\sigma}^{(j)}(t)|^2)\ln (|l_{m'm\sigma}^{(j)}(t)|^2+|l_{m'+1,m\sigma}^{(j)}(t)|^2)}-\left(\sum\limits_{m'=0}^{-\infty}|l_{m'm\sigma}^{(j)}(t)|^2\right)\ln \left(\sum\limits_{m'=0}^{-\infty}|l_{m'm\sigma}^{(j)}(t)|^2\right)\nonumber\\
&-&\left(\sum\limits_{m'=1}^{\infty}|l_{m'm\sigma}^{(j)}(t)|^2\right)\ln \left(\sum\limits_{m'=1}^{\infty}|l_{m'm\sigma}^{(j)}(t)|^2\right)\geq -\sum\limits_{m'=-\infty}^{\infty}|l_{m'm\sigma}^{(j)}(t)|^2\ln |l_{m'm\sigma}^{(j)}(t)|^2
   \end{eqnarray*}
      \end{widetext}
for the indexes $m'\in\{0,\pm1,\pm2,\ldots\}$ and
\begin{widetext}
       \begin{eqnarray*}\label{31}
       &-&{\sum\limits_{m'=-\frac{1}{2}}^{-\infty}(|l_{m'm\sigma}^{(j)}(t)|^2+|l_{m'+1,m\sigma}^{(j)}(t)|^2)\ln (|l_{m'm\sigma}^{(j)}(t)|^2+|l_{m'+1,m\sigma}^{(j)}(t)|^2)}-\left(\sum\limits_{m'=-\frac{1}{2}}^{-\infty}|l_{m'm\sigma}^{(j)}(t)|^2\right)\ln \left(\sum\limits_{m'=0}^{-\infty}|l_{m'm\sigma}^{(j)}(t)|^2\right)\nonumber\\
&-&\left(\sum\limits_{m'=\frac{1}{2}}^{\infty}|l_{m'm\sigma}^{(j)}(t)|^2\right)\ln \left(\sum\limits_{m'=\frac{1}{2}}^{\infty}|l_{m'm\sigma}^{(j)}(t)|^2\right)\geq -\sum\limits_{m'=-\infty}^{\infty}|l_{m'm\sigma}^{(j)}(t)|^2\ln |l_{m'm\sigma}^{(j)}(t)|^2
   \end{eqnarray*}
      \end{widetext}
for the indexes $m'\in\{\pm1/2,\pm3/2,\ldots\}$.
If we substitute their the polynomials \eqref{32} we also can get the new inequalities for the Gauss' hypergeometric functions.

\section{Summary}

To conclude we point out the main results of the work.
Considering the matrix elements of the unitary irreducible
representations of the groups $SU(2)$ and $SU(1,1)$ and applying known subadditivity condition for joint probability distributions constructed from these matrix elements we obtained new inequalities for the Jacobi and the Gauss' hypergeometric polynomials. The inequalities correspond to entropic inequalities for Shannon entropies
of bipartite classical systems. The results are illustrated by the examples of the systems with the spins $j=3/2$ and $j=2$, where the Shannon  information of the bipartite system is expressed in terms of the polynomials. It is shown that using another mappings and entropies, i.e. Tsallis entropy, many other inequalities for the special functions can be written.

\nocite{*}
\bibliographystyle{plain}
\bibliography{aipsamp}

\end{document}